%% file: EMAP.tex
\documentclass[conference,10pt]{IEEEtran}
\IEEEoverridecommandlockouts

\usepackage{booktabs} 
\usepackage{url} 
\usepackage{graphicx} 
\usepackage{amsmath,amssymb,amsfonts} 
\usepackage{xcolor} 
\usepackage[labelfont=bf]{caption} 
\usepackage{textcomp} 
\usepackage{dblfloatfix}
\usepackage{fancyhdr}

\usepackage{cite}
\usepackage{soul} 
\usepackage{enumitem} 
\usepackage{lipsum} 
\usepackage[hang,flushmargin]{footmisc}
\usepackage{etoolbox}
\usepackage{float}
\usepackage{multirow}
\usepackage{hhline}
\newlist{inlinelist}{enumerate*}{1}
\setlist[inlinelist]{label=(\roman*)}

\usepackage{algorithmicx}
\usepackage{algorithm}
\usepackage{algpseudocode}
\usepackage{wrapfig}

\algnewcommand\algorithmicinput{\textbf{Input:}}
\algnewcommand\Input{\item[\algorithmicinput]}

\algnewcommand\algorithmicconst{\textbf{Constraints:}}
\algnewcommand\Const{\item[\algorithmicconst]}

\algnewcommand\algorithmicoutput{\textbf{Output:}}
\algnewcommand\Output{\item[\algorithmicoutput]}

\algnewcommand{\algorithmicgoto}{\textbf{go to}}%
\algnewcommand{\Goto}[1]{\algorithmicgoto~\ref{#1}}%

\algrenewcommand\algorithmicindent{0.5em}

\renewcommand{\thefigure}{\arabic{figure}}

\usepackage{array}
\newcolumntype{L}[1]{>{\raggedright\let\newline\\\arraybackslash\hspace{0pt}}m{#1}}
\newcolumntype{C}[1]{>{\centering\let\newline\\\arraybackslash\hspace{0pt}}m{#1}}
\newcolumntype{R}[1]{>{\raggedleft\let\newline\\\arraybackslash\hspace{0pt}}m{#1}}
\newcolumntype{M}[1]{>{\centering\arraybackslash}m{#1}}
\newcolumntype{O}[1]{>{\raggedleft\arraybackslash}m{#1}}
\usepackage{makecell}
\usepackage{diagbox}

\usepackage[hidelinks, bookmarks=false]{hyperref}

\def\BibTeX{{\rm B\kern-.05em{\sc i\kern-.025em b}\kern-.08em
    T\kern-.1667em\lower.7ex\hbox{E}\kern-.125emX}}
\begin{document}

\title{\LARGE EMAP: A Cloud-Edge Hybrid Framework for \underline{E}EG \underline{M}onitoring and Cross-Correlation Based Real-time \underline{A}nomaly \underline{P}rediction}

\author{
    \IEEEauthorblockN{
    Bharath Srinivas Prabakaran\IEEEauthorrefmark{1}\textsuperscript{,}\IEEEauthorrefmark{3}\thanks{\IEEEauthorrefmark{3}These two authors have contributed to this work equally.}, Alberto Garc\'{i}a Jim\'{e}nez\IEEEauthorrefmark{1}\textsuperscript{,}\IEEEauthorrefmark{2}\textsuperscript{,}\IEEEauthorrefmark{3}, Germ\'{a}n Molt\'{o} Mart\'{i}nez\IEEEauthorrefmark{2}, Muhammad Shafique\IEEEauthorrefmark{1}}
    \IEEEauthorblockA{\IEEEauthorrefmark{1}Institute of Computer Engineering, Technische Universit{\"a}t Wien (TU Wien), Austria
    \\\{bharath.prabakaran, muhammad.shafique\}@tuwien.ac.at}
    \IEEEauthorblockA{\IEEEauthorrefmark{2}Sistemas Inform\'{a}ticos y computaci\'{o}n, Universitat Polit\'{e}cnica de Val\'{e}ncia, Spain
    \\\{algarji1@etsid.upv.es, gmolto@dsic.upv.es\}}
}


\fancypagestyle{firstpage}{
  \fancyhf{}
  \fancyhead[C]{To appear at the 57th Design Automation Conference (DAC), July 2020, San Francisco, CA, USA.}
  \fancyfoot[C]{\thepage}
}

\maketitle

\thispagestyle{firstpage}
\pagestyle{plain}

\begin{abstract}
State-of-the-art techniques for detecting, or predicting, neurological disorders (1) focus on predicting each disorder individually, and are (2) computationally expensive, leading to a delay that can potentially render the prediction useless, especially in critical events.
Towards this, we present a real-time two-tiered framework called EMAP, which cross-correlates the input with all the EEG signals in our mega-database (a combination of multiple EEG datasets) at the cloud, while tracking the signal in real-time at the edge, to predict the occurrence of a neurological anomaly. 
Using the proposed framework, we have demonstrated a prediction accuracy of up to 94\% for the three different anomalies that we have tested.
\end{abstract}

\bstctlcite{IEEEexample:BSTcontrol}

\input{sections/section1.tex}
\input{sections/section2.tex}
\input{sections/section3.tex}
\input{sections/section4.tex}
\input{sections/section5.tex}
\input{sections/section6.tex}
\input{sections/section7.tex}

\section*{Acknowledgement} 
This work was partially supported by Doctoral College Resilient Embedded Systems which is run jointly by TU Wien's Faculty of Informatics and FH-Technikum Wien.

\bibliographystyle{IEEEtran}
\bibliography{References}

\end{document}

%% file: sections/section1.tex
\section{Introduction}
\label{sec:Introduction}

The human brain is responsible for performing a wide range of autonomous, semi-autonomous, and manual functions, such as generating thoughts, motor control, storing memories, regulating hormones, etc.~\cite{herculano2009human}.
It is susceptible to more than 600 diseases like tumors, epilepsy, Alzheimer's, and strokes~\cite{WHO2006}.
These diseases can be diagnosed using medical imaging techniques, such as (functional) Magnetic Resonance Imaging (fMRI/MRI) and Computed Tomography (CT Scan), and Electroencephalography (EEG)~\cite{AANS}.
These techniques are typically used to study the brain \textit{ex-post-facto}, i.e., after the event has occurred, to evaluate the amount of damage that has been caused by the event~\cite{NINDS2019}. 
Patients prone to such neurological orders could be in potentially fatal situations, where they could place themselves and other people in harm, for example, the occurrence of seizures or strokes in drivers and heavy equipment users.

State-of-the-art techniques, typically, analyze the EEG signals using compute-intensive algorithms and statistical methods, including machine learning/deep neural networks, to accurately detect/predict each neurological disorder \textit{individually}~\cite{akmandor2017keep,mardi2011eeg,burrello2019laelaps,pascual2019self,hussein2018epileptic,zhang2018integration,hosseini2016cloud,kim2018wave2vec,samie2018highly}.
This requires the edge device (typically a wearable or a sensor-head) to continuously transmit the EEG data to the cloud for further processing and feature extraction.
Besides the communication time overheads, these fully cloud-based techniques pose serious privacy and security concerns for the users who might not wish to continuously transmit all of their bio-signal data over an insecure/untrustworthy network, or to store it on the third-party cloud platforms~\cite{shirazi2017extended}.
However, it may still be feasible to transmit certain parts of the data to the cloud, if extensive processing is required to recover from life-threatening situations, considering the fact that the third party cannot retrieve the complete signal information with incomplete data.
Such a situation is more realistic and can be considered as a trade-off between privacy, security, and urgency-of-extensive-analysis.

Enabling such an efficient EEG processing system requires addressing the following \textbf{\textit{research challenges}}, as addressed in this work:
\begin{enumerate}[leftmargin=*,label=(\arabic*)]
    \item How can the continuous monitoring of EEG signals at the edge device be used to predict multiple different neurological anomalies?
    \item How to enable real-time anomaly prediction with the help of a cloud-edge hybrid framework, while trying to minimize the amount of data transmitted to the cloud?
\end{enumerate}

\textbf{Novel Contributions:} To address the above challenges, we propose the novel EMAP framework for predicting anomalies in real-time that employs the following key components:
\begin{itemize}[leftmargin=*]
    \item An efficient edge sensor node to continuously monitor the brain signals by collecting, pre-processing, and transmitting only one second of the EEG signal data to the cloud every few seconds;
    \item A Mega-database ($MDB$) of EEG signals, on the cloud, that was constructed by combining various state-of-the-art EEG databases containing normal and anomalous EEG signals, such as seizures, epilepsy, etc.;
    \item A novel signal cross-correlation search algorithm, which efficiently compares the patient's one-second EEG signal with all the signals of the $MDB$ in the cloud, to quickly identify the top-$100$ analogous signals, which are transmitted to the edge;
    \item A novel real-time signal tracking algorithm at the edge to estimate the similarity of the top-$100$ analogous signals with the input in real-time, to eliminate dissimilar signals, estimate the probability of an anomaly, and predict its occurrence based on the inputs obtained from the subsequent time-steps.
\end{itemize}
Furthermore, to enable an efficient design of the EMAP framework, we perform a motivational analysis that studies the benefits of continuous monitoring and signal cross-correlation to estimate the probability of anomalies.
The prediction accuracy of EMAP is evaluated for three different neurological disorders, namely, seizures, strokes, and encephalopathy, using $100$ different input signals for each disorder.

\textbf{Evaluation \& Open-Sourcing:} We have successfully obtained a prediction accuracy of 94\%, 73\%, and 79\%, on average, for the three different anomalies that we have tested.
To enable easy reproduction and adaption of the proposed EMAP framework, we will open-source the complete tool-flow at \textcolor{blue}{\url{https://emap.sourceforge.io}}.
Fig.~\ref{fig:UCDiagram} illustrates an overview of the contributions (in dark highlighted blocks) that have been proposed in the cloud-edge hybrid framework.

\setcounter{figure}{0}
\begin{figure}[h]
    \centering
    \captionsetup{singlelinecheck=false}
    \includegraphics[width = \linewidth]{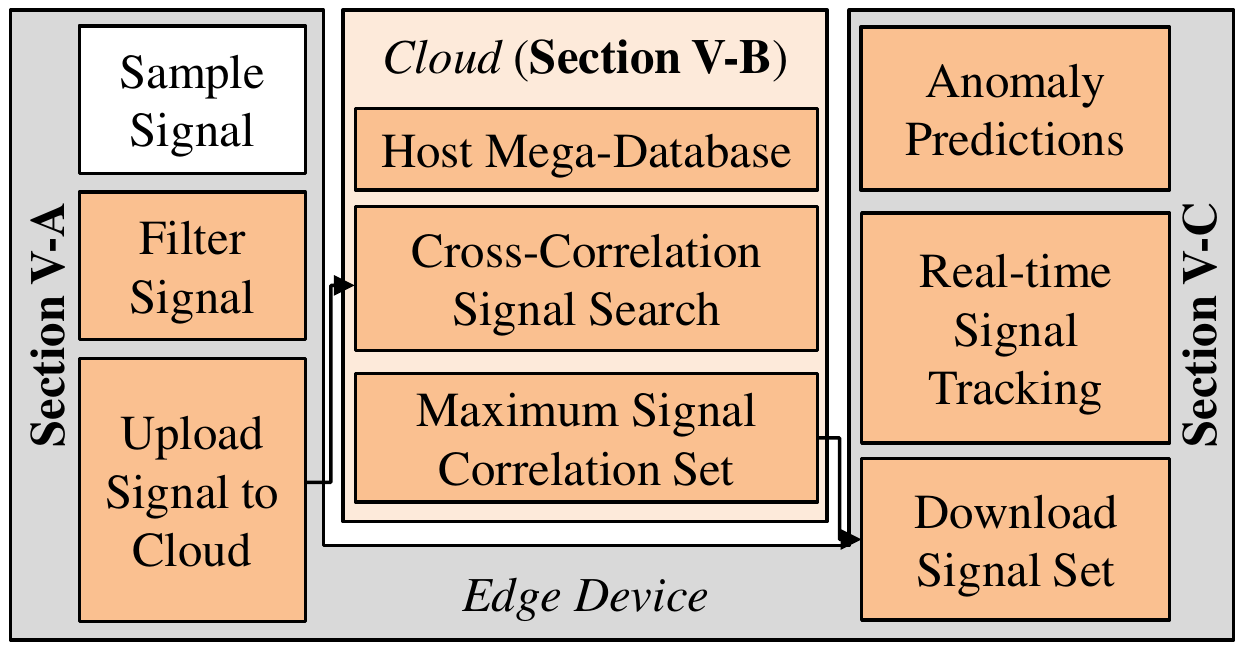}
    \caption{\textbf{Overview of the proposed contributions (dark highlighted boxes) in the cloud-edge hybrid framework.}}
    \label{fig:UCDiagram}
\end{figure}

%% file: sections/section2.tex
\section{Related Work}
\label{sec:RelatedWork}

Electroencephalography (EEG) is a technique that is typically used to study the brain \textit{ex-post-facto}, by medical experts, to ascertain the amount of damage caused due to a specific event.
This typically requires high-quality EEG electrodes that are not portable or easy to use and can be highly expensive.
For the purpose of continuous monitoring, 10-20 electrodes (considered to be an EEG placement standard), which cover the surface area of the head, can be placed on a cap (electrode-caps) to measure the EEG signal samples~\cite{OpenBCI}.
These devices can be used for other purposes, such as stress detection and mitigation~\cite{akmandor2017keep} and drowsiness detection~\cite{mardi2011eeg}.

Recently, these devices are used as wearables in the healthcare industry to accurately detect or predict specific brain anomalies, especially seizures.
Most of the current techniques heavily rely on deep learning for accurate \textit{seizure detection}.
Burrello \textit{et al.} proposed a hyperdimensional computing approach called \textit{Laelaps} that can be used for accurately classifying seizures using EEG signals~\cite{burrello2019laelaps}.
Pascual \textit{et al.} proposed a minimally supervised algorithm that can be used to automatically label seizures, without the help of medical experts, to generate personalized training data~\cite{pascual2019self}.
Similarly, other deep learning techniques have been proposed for seizure detection by Zhou \textit{et al.}~\cite{zhou2018epileptic} and Hussein \textit{et al.}~\cite{hussein2018epileptic}.
On the other hand, various research works have also addressed the problem of \textit{seizure prediction} using deep learning by Zhang \textit{et al.}~\cite{zhang2018integration}, Hosseini \textit{et al.}\cite{hosseini2016cloud}, and Kim \textit{et al.}~\cite{kim2018wave2vec}.
Typically, these techniques are resource-consuming and might require additional hardware units such as powerful deep neural network accelerators or GPUs, which may not be feasible for low-cost IoT edge devices.
Towards this, Samie \textit{et al.} proposed an algorithm for seizure prediction that can be deployed on low-power resource-constrained IoT devices~\cite{samie2018highly}.
Previous research works have also proposed the use of signal cross-correlation for diagnosing physical and psychological diseases such as epilepsy and schizophrenia~\cite{timashev2012analysis}\cite{zhang2015seizure}.

In this work, a \textit{cloud-edge hybrid framework} has been proposed, which can monitor EEG signals and \textit{predict the occurrences of various brain-related anomalies}, and not just seizures, \textit{in real-time}.

%% file: sections/section3.tex
\section{Preliminaries}
\label{sec:Preliminaries}

In this section, we present the relevant background knowledge to a level of detail that is necessary to understand the proposed novel contributions in this work.

\textbf{Bandpass Filters:}
Finite Impulse Response (FIR) Bandpass filters are used to attenuate the noise components and motion artifacts outside the desired frequency range.
This is especially the case for multi-channel EEG electrodes that are highly susceptible to noise because of the location of their placement, i.e., on the scalp of the users.
Therefore, we define a 100-tap FIR bandpass filter ($H(z)$), which attenuates all frequencies besides the desired range of $11-40$ Hz, with the following transfer function:

\begin{equation}
    H(z) = \sum_{n=0}^{99}h(n)z^{-n}
\end{equation}

\textbf{Signal Cross-Correlation:}
The similarity of two signals can be evaluated using a metric known as cross-correlation, which is a function of the displacement of one signal with respect to the other, also known as the sliding dot product.
The cross-correlation of two signals $A_N$ and $B_M$, composed of 256 samples each, is defined as:

\begin{equation}
    \omega(A_N,B_N) = \sum_{n=0}^{n=255}A_{(N,n)} \text{x} B_{(M,n)}
\end{equation}

\textbf{Area Between Curves:}
Besides cross-correlation, the similarity of two signals can also be determined by calculating the area between the curves of the two signals, which is defined as:

\begin{equation}
    A(A_N,B_M)= \sum_{i=0}^{i=255} |A_{(N,i)} - B_{(M,i)}|
\end{equation}

\textbf{Time Consumption:}
The initial time overhead ($\Delta_{initial}$) for the proposed framework to estimate anomaly probabilities for the first iteration after deployment is modeled using the following equation:

\begin{equation}
    \Delta_{initial} = \Delta_{EC} + \Delta_{CS} + \Delta_{CE}
\end{equation}

where $\Delta_{EC}$ is the time required for transmitting the input EEG signal from the edge to the cloud, $\Delta_{CS}$ is the time required for the signal cross-correlation search to determine the set of signals with maximum similarity to the input signal, i.e., cloud search, and $\Delta_{CE}$ is the time required to download this set of signals, from the cloud, by the edge device.
In each subsequent time-step, the signal tracking algorithm at the edge is used to estimate the anomaly probabilities, which is required to be less than one second for the proposed EMAP framework to meet the real-time constraints.

\textbf{Anomaly Probability:}
We define the probability of occurrence of an anomaly ($P_A$) as the proportion of anomalous signals ($N(A_S)$) with respect to the total number of signals being tracked ($N(F)$) at the edge. It can be computed as:

\begin{equation}
    P_A = N(A_S)/N(F)
\end{equation}

%% file: sections/section4.tex
\section{Analysis of Signal Cross-Correlation for Predicting Brain Anomalies}
\label{sec:Motivation}

To illustrate that cross-correlation can be used to predict anomalies, we performed an experiment to determine the top-$100$ signals in the $MDB$ with maximum similarity to an anomalous input signal.
As Fig.~\ref{fig:CCAP}(a) illustrates, the proportion of normal to anomalous signals is quite large, and if its probability were to be estimated at this point, it would be very low, i.e., close to $22\%$.
However, with continuous monitoring, dissimilar signals can be eliminated in real-time to only keep track of signals that are highly correlated to the input signal; see Figs.~\ref{fig:CCAP}(b)-(f).
Using this approach, we eliminate dissimilar signals after each time-step/iteration and estimate the probability of anomalies, which goes up to $66\%$ at the end of \textit{Iteration 5}.
The probability of occurrence of an anomaly increases after each iteration as the proposed EMAP framework eliminates the normal signals at a higher rate as compared to the anomalous signals, due to their dissimilarity to the input signal.

\setcounter{figure}{1}
\begin{figure}[t]
    \centering
    \captionsetup{singlelinecheck=false}
    \includegraphics[width = \linewidth]{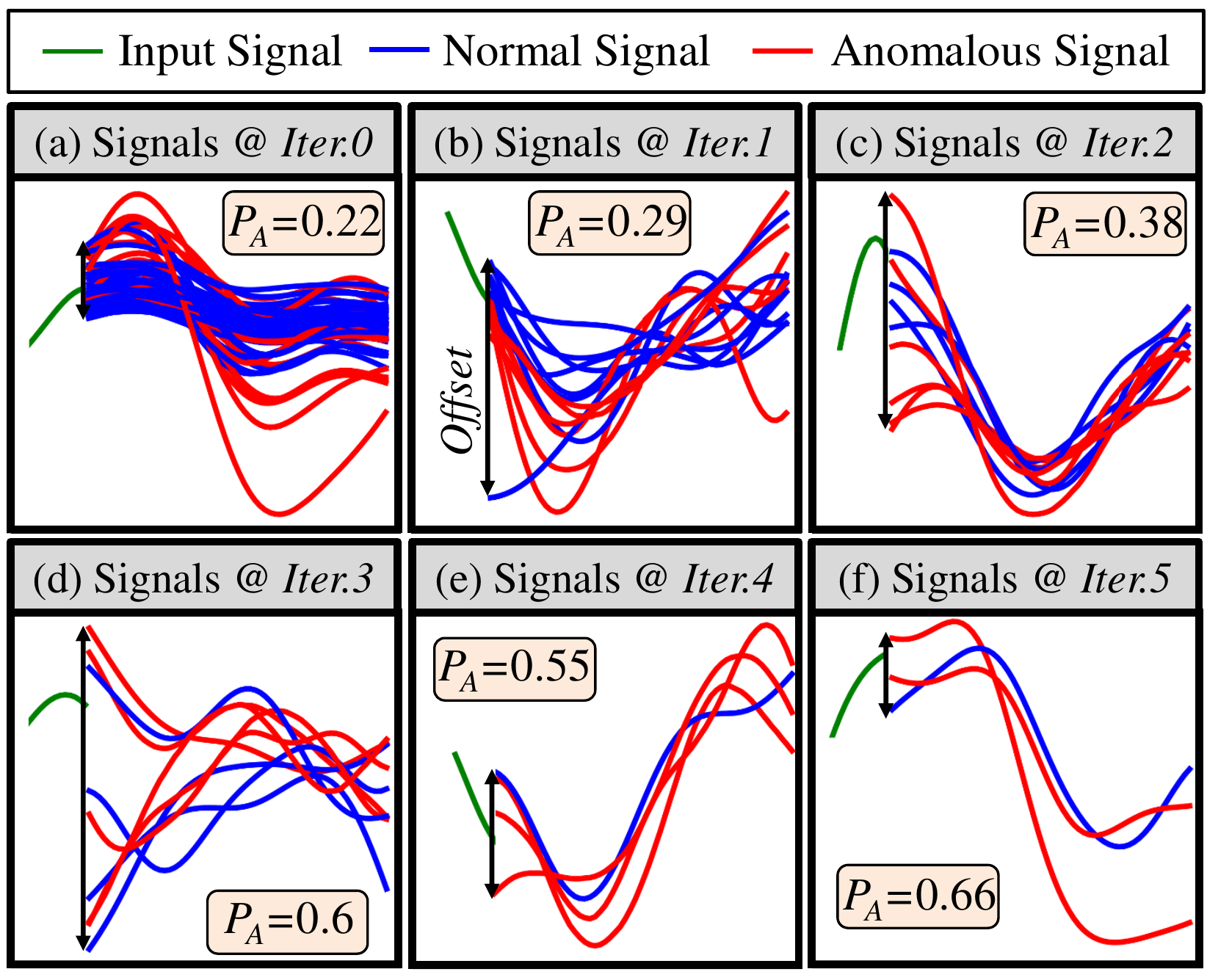}
    \caption{\textbf{Analysis for the use of cross-correlation on EEG signals for anomaly prediction.}{ \normalfont Tracking the number of normal and anomalous signals with high correlation to the input, at each iteration (each second), to estimate anomaly probability ($P_A$).}}
    \label{fig:CCAP}
\end{figure}

\setcounter{figure}{2}
\begin{figure*}[t]
    \centering
    \captionsetup{singlelinecheck=false}
    \includegraphics[width = \linewidth]{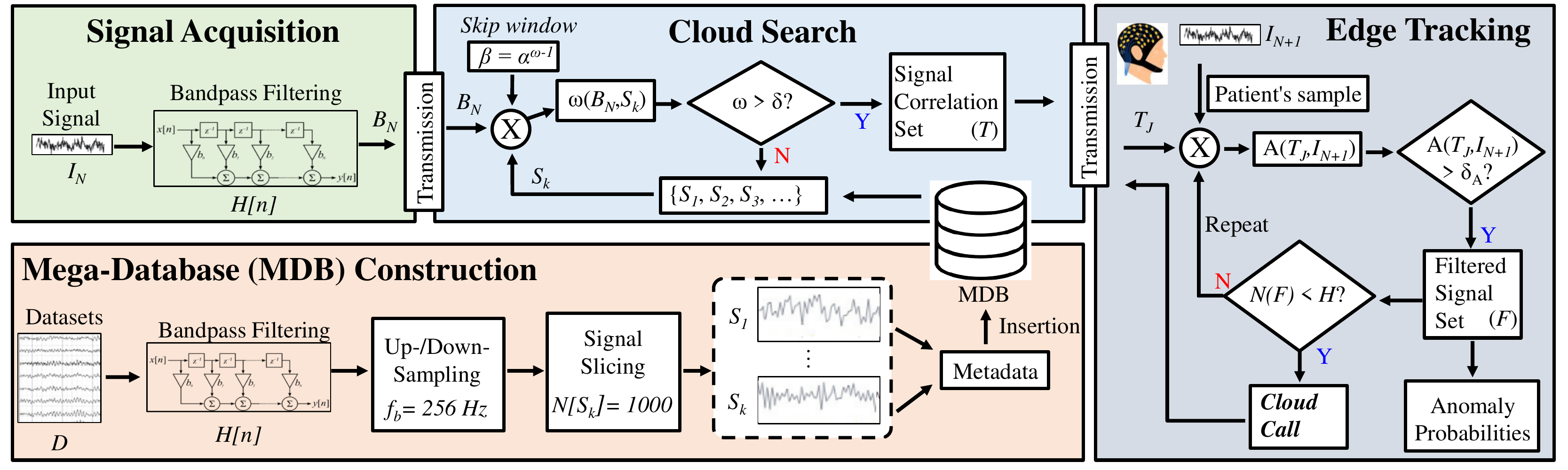}
    \caption{\textbf{An overview of the proposed EMAP framework.}{ \normalfont Three key stages: (a) Signal Acquisition at the edge, which pre-processes the input ($I_N$) and transmits data to the cloud; (b) Cloud Search, where the input ($B_N$) is cross-correlated with all signals in the $MDB$ to identify top-$100$ maximally correlated signals ($T$) that are transmitted to the edge; and (c) Edge Tracking, which evaluates the similarity between the input signal ($I_{N+1}$) and the correlated set of signals ($T$) to enable real-time tracking.}}
    \label{fig:EMAP}
\end{figure*}

\setcounter{figure}{3}
\begin{figure}[t]
    \centering
    \captionsetup{singlelinecheck=false}
    \includegraphics[width = \linewidth]{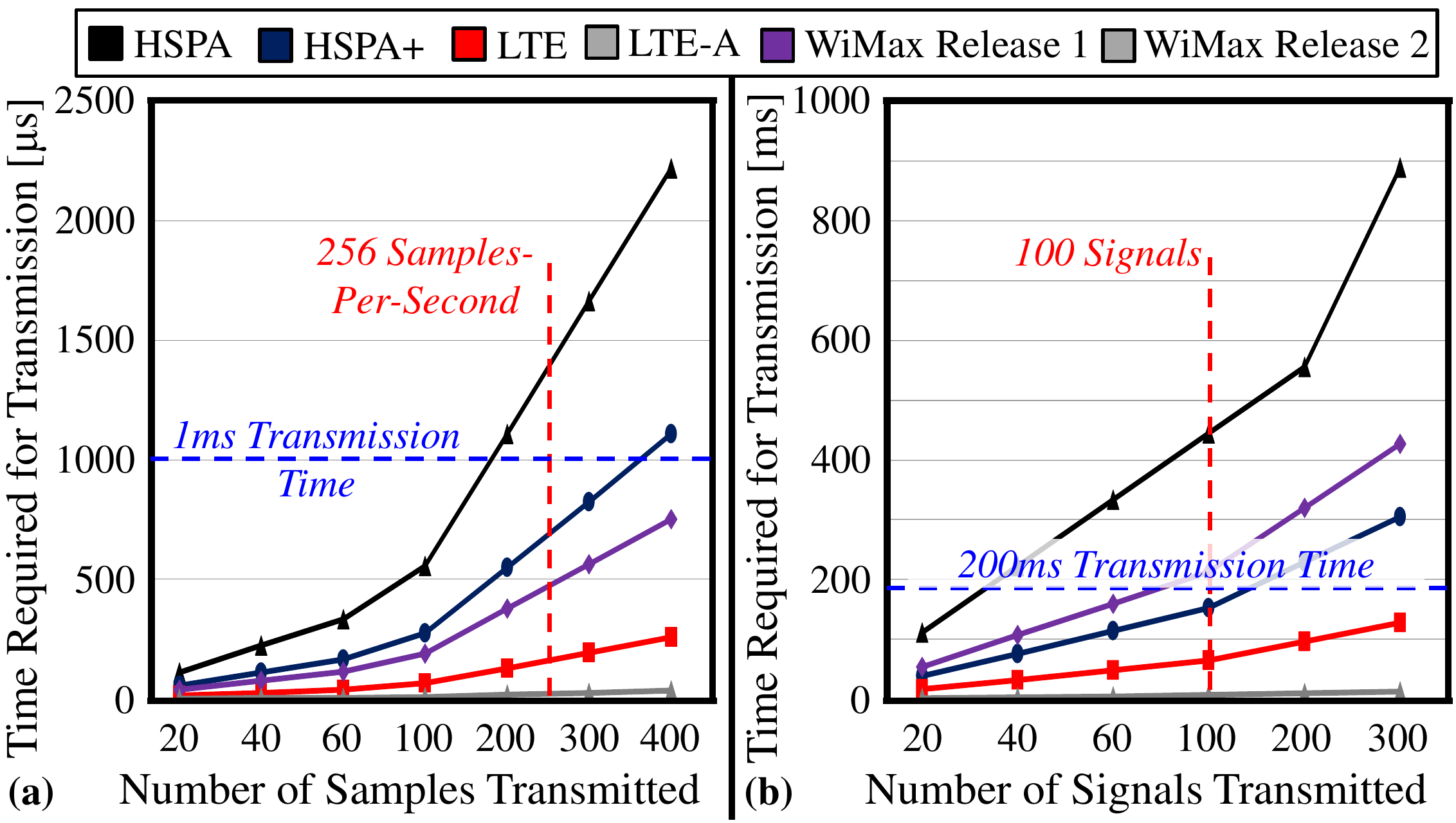}
    \caption{\textbf{Time required for upload [in $\mu s$] and download [in $ms$] across different communication platforms for varying number of samples and signals (adapted from data presented in~\cite{steer2007beyond}\cite{parkvall2008lte}).}}
    \label{fig:Communication}
\end{figure}

%% file: sections/section5.tex
\section{Our EMAP Framework}
\label{sec:EMAP}

Fig.~\ref{fig:EMAP} presents an overview of the proposed EMAP framework, which is composed of three key stages that are explained in detail in the subsequent sub-sections:
\begin{enumerate}[leftmargin=*,label={(\arabic*)}]
    \item \textit{Signal Acquisition} is responsible for sampling, pre-processing, and transmitting one-second of EEG signal data to the cloud,
    \item \textit{Cloud Search} compares the input signal against all the signals in the $MDB$ using the proposed cross-correlation search algorithm to identify the top-$100$ analogous signals, which are transmitted to the edge for real-time tracking, and
    \item \textit{Edge Tracking} employs a novel signal-tracking algorithm to evaluate the similarity of the incoming input samples with the set of analogous signals to eliminate dissimilar signals and estimate the probability of occurrence of an anomaly in real-time.
\end{enumerate}
The cloud-edge hybrid framework allows us to effectively offload the compute- and memory-intensive signal cross-correlation algorithm to the cloud while the searched signals are tracked, and their anomaly probability estimated, in real-time by the edge device.

\subsection{Signal Acquisition}
This stage of the framework is responsible for three main tasks, namely,
\begin{inlinelist}
\item signal sampling,
\item filtering, and
\item transmission.
\end{inlinelist}
The electrode-caps, discussed in Section~\ref{sec:RelatedWork}, can be used as sensor nodes to sample the brain signals at the required frequency of $256$ Hz (16-bit resolution), after which the signal is filtered using a bandpass filter to eliminate the noise components and to generate a uniform piece-wise linear curve that can be transmitted to the cloud for comparison.
We define the input signal obtained from the EEG headset as $I_N = \{I_{(N,1)}, I_{(N,2)}, ... I_{(N,256)}\}$, where $I_N$ denotes the set of samples of the input signal at time-step $N$, and $I_{(N,k)}$ denotes the $k^{th}$ sample in the $N^{th}$ time-step of the input.
This signal is passed through the $100$-tap bandpass filter to obtain the signal $B_{(N,k)} = \sum_{i=0}^{i=99} H_i \times I_{(N,k-i)}$, which is subsequently transmitted to the cloud.
Note, it might be suitable to include a 100-tap bandpass filter as a simple hard-coded accelerator on the edge device, to ensure that the framework works in real-time.
Fig.~\ref{fig:Communication}(a) presents the time required for uploading the various number of samples to the cloud across different communication platforms.
In the era of 4G communication, the time required for transmitting an EEG signal of one time-step should ideally take less than $1$ms. 
This time constraint is imposed to efficiently offload compute- and memory-intensive tasks to the cloud, and receive their outputs in real-time.

\subsection{Cloud Search}
This stage is composed of two parts, namely,
\begin{inlinelist}
\item the construction of the $MDB$, and
\item the signal cross-correlation search.
\end{inlinelist} 
The first part, i.e., the construction of the $MDB$, involves identifying and combining $5$ different state-of-the-art open-access EEG datasets presented in~\cite{PhysioNet}\cite{harati2014tuh}\cite{Dua:2019}\cite{brunner2014bnci}\cite{zwolinski2010open}, to include a wide-range of normal and anomalous EEG waveforms.
This includes collecting, up-/down-sampling the signals to the base frequency of $256$ Hz, and labeling the EEG signals as normal or anomalous.
Since the input signal to this stage is bandpass filtered, all the signals in the dataset are also bandpass filtered to ensure consistency, uniformity, and ease of search.
Each signal in the dataset is further sliced into signal-sets of $1000$ samples each, and allocated a label (normal or anomalous), to enable the search algorithm to quickly search through the complete database in parallel.
We define the super-set of datasets as $D = \{D_1, D_2, ... D_X\}$, where each dataset $D_W$ is composed of a set of signals $L(D_W) = \{L_1, L_2, ...\}$, such that $W \in \{1, 2, ...X\}$.
Each signal, in each dataset, is passed through the bandpass filter to generate the output signal $BL_{Y,(N,k)} = \sum_{i=0}^{i=99} H_i \times L_{Y,(N,k-i)}$.
Each signal $BL_Y$ is sliced into signal-sets of $1000$ samples each, which are then subsequently labeled as normal or anomalous to construct the mega-database $MDB$.
The $MDB$ is defined as the super-set of all signal-sets, i.e., $\{S_1, S_2, S_3, ...\}$, where each signal-set has the attribute/label $A(S_P)$, such that $A(S_P)=0$ for normal signals and $A(S_P)=1$ for anomalous signals.

\begin{algorithm}[b]
    \normalsize
    \caption{The Signal Cross-Correlation Search}
    \label{Algo1}
    \begin{algorithmic}[1]
        \Input $\text{Input Signal} (I_N), \text{Mega-Database} (MDB)$
        \Const $\text{Step-size} (\alpha), \text{Cross-Correlation Threshold} (\delta)$
        \Output $\text{Signal Correlation Set} (T)$
        \State $SignalArray = Array[];$
        \For{$S$ \textbf{in} $MDB$}
        \State $\beta = 0;$
        \While{$\beta < Length(S) - Length(I_N)$}
        \State $\omega = \sum_{n=0}^{n=255} I_{(N,n)} \times S_{(\beta:\beta+Length(I_N),n)};$
        \If{$\omega > \delta$}
        \State $SignalArray.append([S,\omega, \beta]);$
        \EndIf
        \If{$\omega < 0$}
        \State $\omega = 0;$
        \EndIf
        \State $\beta = \beta + \alpha^{\omega-1}$
        \EndWhile
        \EndFor
        \State $AscendingSort(SignalArray, \omega);$
        \State $T=SignalArray(0:99);$
    \end{algorithmic}
\end{algorithm}

\begin{figure}[t]
    \centering
    \captionsetup{singlelinecheck=false}
    \includegraphics[width = 0.92\linewidth]{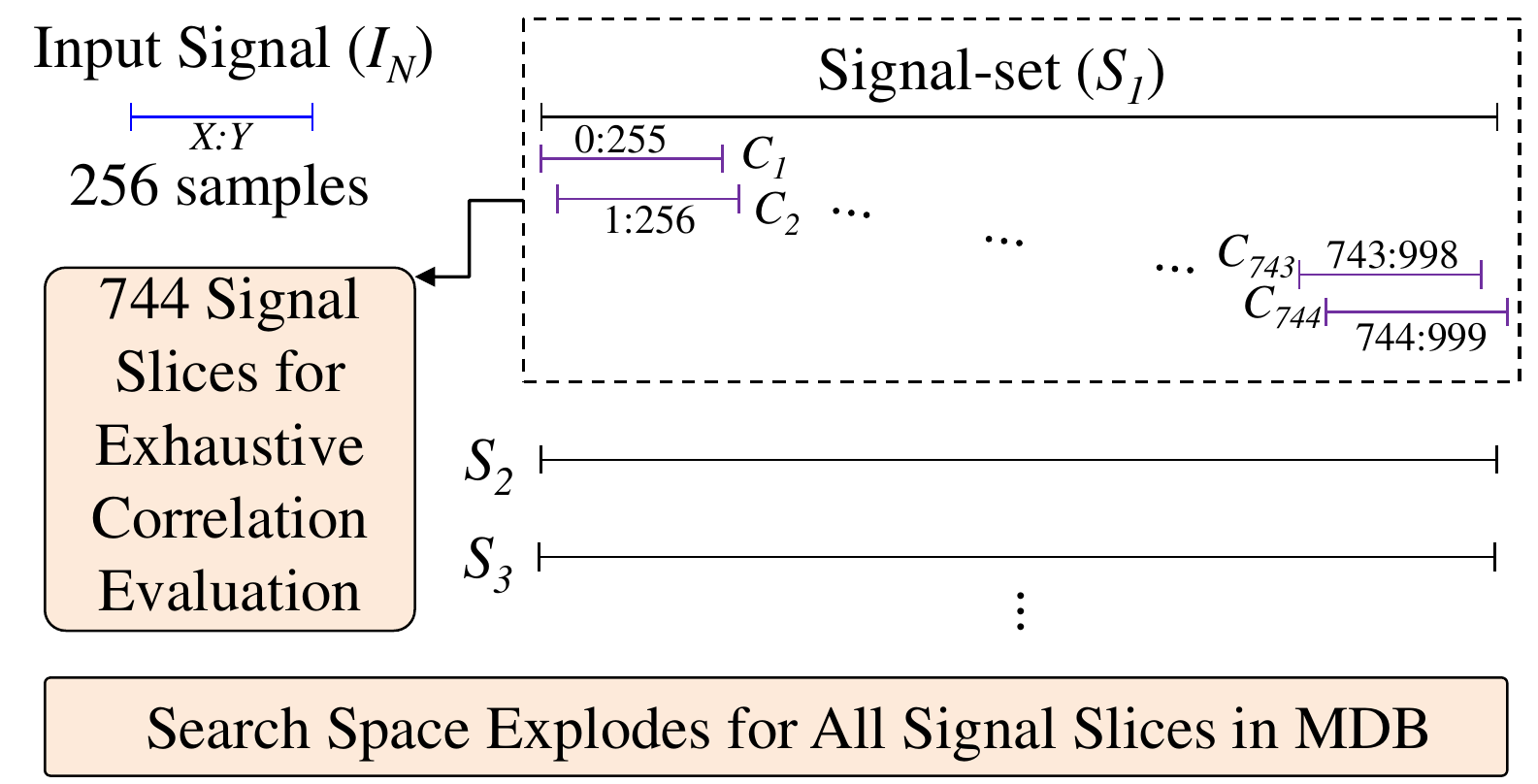}
    \caption{\textbf{Illustration of the search space explosion for exhaustive cross-correlation of the input signal and all slices in the MDB.} The length of a slice is denoted in the [\textit{X}:\textit{Y}] format, where \textit{X} is the starting number of the sample and \textit{Y} is the ending number of the sample.}
    \label{fig:StepSize}
\end{figure}

\begin{figure}[t]
    \centering
    \captionsetup{singlelinecheck=false}
    \includegraphics[width = 0.92\linewidth]{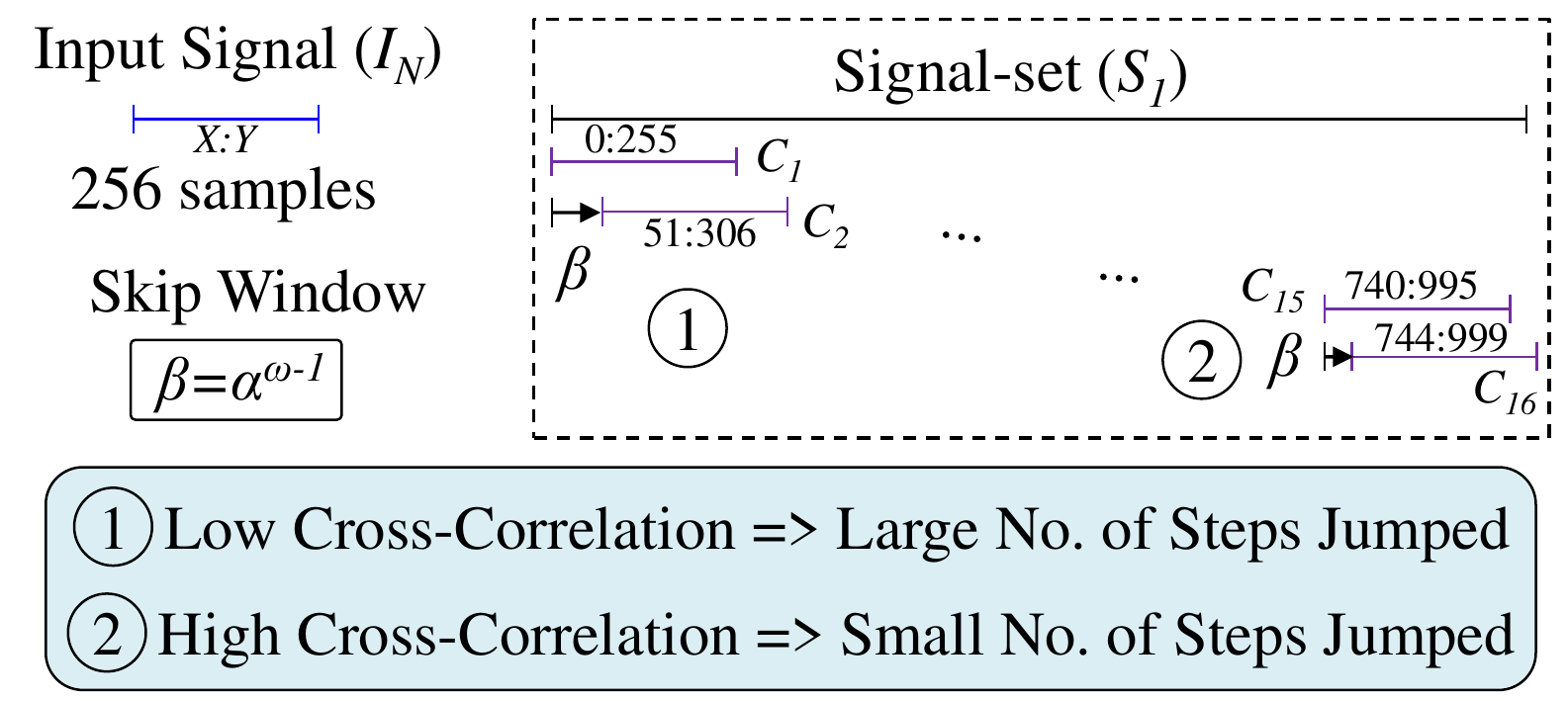}
    \caption{\textbf{Illustration of the proposed sliding window technique to reduce the search space.} The length of a slice is denoted in the [\textit{X}:\textit{Y}] format, where \textit{X} is the starting number of the sample and \textit{Y} is the ending number of the sample.}
    \label{fig:StepSizeProp}
\end{figure}

\begin{figure}[t]
    \centering
    \captionsetup{singlelinecheck=false}
    \includegraphics[width = \linewidth]{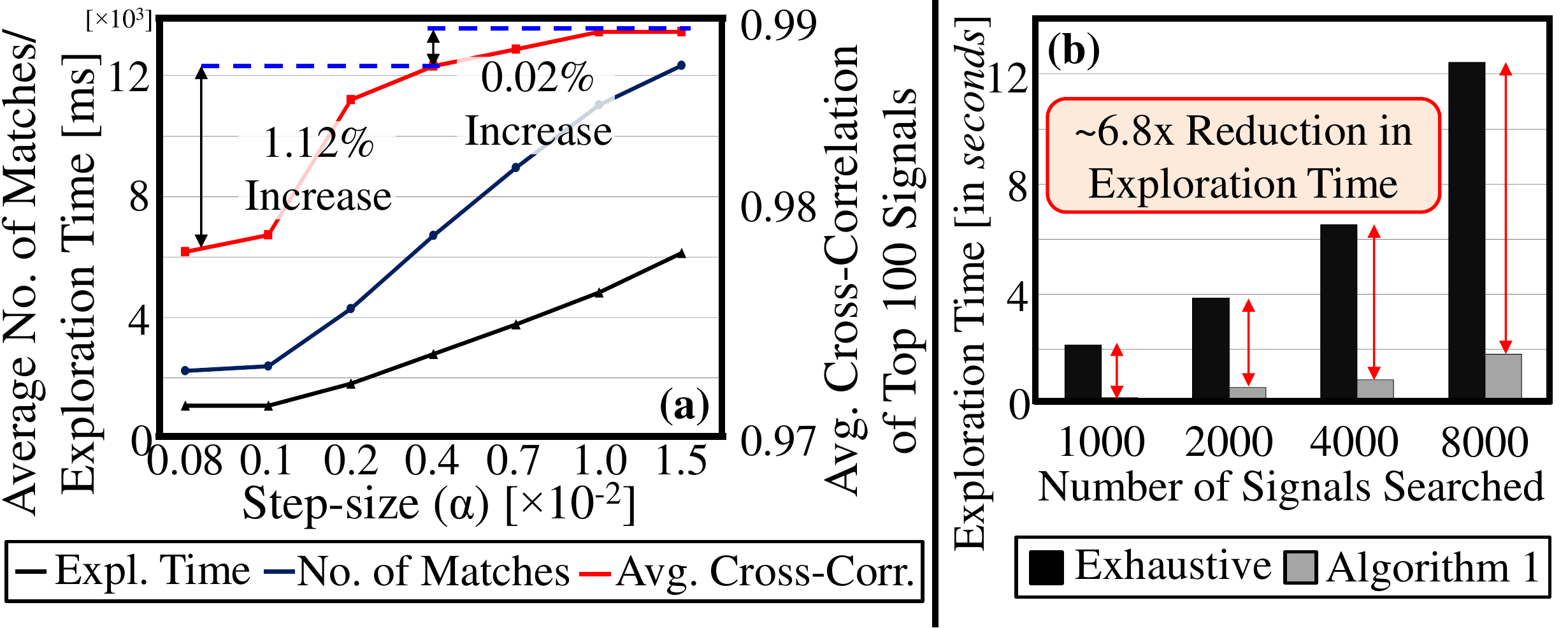}
    \caption{\textbf{(a) Analysis of the average number of matches, exploration time, and average cross-correlation of the top-$100$ Signals for varying values of step-size ($\alpha$); (b) exploration time for exhaustive search \& Algorithm~\ref{Algo1}.}}
    \label{fig:EvolutionAlpha}
\end{figure}

The next part in this stage cross-correlates the input signal $I_N$ obtained from the signal acquisition stage with all the signal-sets present in the $MDB$.
Because of the large number of signal-sets, the search space for this algorithm is huge; therefore, the time required for this stage is also quite large.
For example, an input signal composed of $256$ samples needs to be cross-correlated $744$ times with a single signal-set of $1000$ samples in the $MDB$, with a skip window ($\beta$), i.e., the number of offset samples, of $1$, in each iteration. This example is depicted in Fig.~\ref{fig:StepSize}.
Therefore, we use a sliding window approach that offsets the signal-set, based on the step-size ($\alpha$), before each cross-correlation is performed. 
We propose to increase/decrease the step-size non-linearly because of the following reasons:
\begin{inlinelist}
\item two dissimilar signals can increase the number of searches, without identifying similar signals, when $\alpha$ is very small, and 
\item two very similar signals can reduce the number of searches, by skipping over similar signals, when $\alpha$ is very large.
\end{inlinelist}
Therefore, we define an exponential sliding window, which can increase or decrease the skip window ($\beta$) based on the correlation of the signals obtained at offset $0$ and the step-size, as $\beta = \alpha^{\omega-1}$.
The proposed approach considers the aforementioned properties of EEG signals to reduce the search space, as depicted in Fig.~\ref{fig:StepSizeProp}.
We determine the value of $\alpha$ for the framework by performing a series of experiments to study the exploration time, the number of correlated signal matches, and average cross-correlation in the top-$100$ signals by varying the values of $\alpha$.
The results of these experiments are illustrated in Fig.~\ref{fig:EvolutionAlpha}(a).
As illustrated in the figure, the signal cross-correlation value saturates and increases by very small margins when $\alpha$ is increased beyond $0.004$. 
This ensures that highly correlated signals are not eliminated during the proposed signal cross-correlation search.
Therefore, we have preset $\alpha$ to be $0.004$ for all the future requirements in our proposed framework, to limit the initial overhead ($\Delta_{initial}$) to \texttildelow$3$ seconds.
We also illustrate the benefits of using the proposed approach when compared to the exhaustive search for a varying number of signal-sets explored in Fig.~\ref{fig:EvolutionAlpha}(b).
On average, we achieve \texttildelow$6.8\times$ reduction in the exploration time when using the proposed signal cross-correlation algorithm, i.e., Algorithm~\ref{Algo1}, when compared to the exhaustive search.
The Algorithm searches over the complete signal-set space $MDB$ to determine the set of top-$100$ signals ($T$), which have the maximum correlation with the input signal $I_N$.
Based on our experiments, the cross-correlation threshold $\delta$ was determined to be $0.8$ in order to avoid scenarios where the input signal $I_N$ is unable to find similar signals in the $MDB$.
This set of cross-correlated signals is transmitted to the edge device for real-time signal tracking and anomaly prediction.

\setcounter{figure}{8}
\begin{figure*}[t]
    \centering
    \captionsetup{singlelinecheck=false}
    \includegraphics[width = \linewidth]{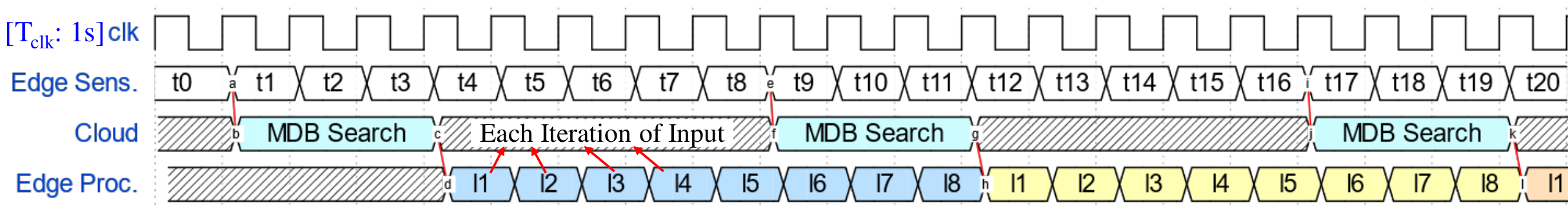}
    \caption{\textbf{Timing Analysis of the Proposed EMAP Framework.}{ \normalfont The sensor node samples the EEG signal at $256$Hz, i.e., $256$ samples for each time-step \{$t0$, $t1$, $t2$, ...\}. The system incurs an initial latency overhead of \texttildelow$3$s caused by the \textit{MDB Search} in the Cloud. After the Cloud Search is complete, the top-$100$ signals are transmitted to the edge for real-time edge tracking at each iteration \{$I1$, $I2$, $I3$, ...\}.}}
    \label{fig:TimingDiagram}
\end{figure*}

\begin{algorithm}[b]
    \normalsize
    \caption{The Lightweight Signal Tracking at the Edge}
    \label{Algo2}
    \begin{algorithmic}[1]
        \Input $\text{Input Signal} (I_{N+1}), \text{Signal Correlation Set} (T)$
        \Const $\text{Threshold for Signal Tracking} (H)$
        \Const $\text{Area Threshold} (\delta_A)$
        \Output $\text{Filtered Signal Set} (F)$
        \While{\texttt{True}}
        \State $F = T;$
        \For{$W$ \textbf{in} $F$}
        \While{$W.\beta < Length(S) - Length(I_{N+1})$}
        \State $A = \sum_{n=0}^{n=255} |I_{(N+1,n)} - W.S_{(\beta:\beta+Length(I_{N+1}),n)}|;$
        \If{$A > \delta_A$}
        \State $F.remove(W);$
        \EndIf
        \EndWhile
        \EndFor
        \If{$Length(F) \leq H$}
        \State $CloudCall(I_{N+1});$
        \EndIf
        \EndWhile
    \end{algorithmic}
\end{algorithm}

\subsection{Edge Tracking}

Fig.~\ref{fig:Communication}(b) presents the time required to download the signal correlation set from the cloud for various values of the number of signals transmitted. 
For the framework to work in real-time, the complete signal correlation set needs to be downloaded in less than $200$ms.
In the \textit{Edge Tracking} stage, we propose a simple lightweight algorithm for tracking the signal using the signal's input from the next time-step, i.e., $N+1$.
Re-evaluating the cross-correlation for each of the $100$ signals is both time and resource consuming, neither of which are, typically, available in the embedded edge nodes that are required to perform computations in real-time and are resource-constrained.
Therefore, we propose to evaluate the area between the curves for the subsequent time-steps to estimate the similarity between the input signal $I_{N+1}$ and the signals present in $T$.
The proposed lightweight signal tracking algorithm is illustrated in Algorithm~\ref{Algo2}.
Next, we compare the two signal matching techniques by varying the cross-correlation ($\delta$) and area thresholds ($\delta_A$) to determine the number of signal matches that can be obtained.
The results of this experiment are presented in Fig.~\ref{fig:EdgeTracking}(a). Based on this analysis, we can determine that the area threshold for the edge tracking algorithm is equal to \texttildelow$900$ sq. units, which is roughly equivalent to the signal cross-correlation threshold ($0.8$) deployed in the cloud.
This threshold value can be modified to increase/decrease the number of matches based on the user requirements and the processing capabilities of the edge device.
Moreover, we have also performed an experiment to determine the execution time differences between the cross-correlation approach and the proposed technique, the results of which are presented in Fig.~\ref{fig:EdgeTracking}(b).
This method of estimating the similarity is roughly \texttildelow$4.3\times$ faster.
Furthermore, the time required for tracking $100$ signals, by the edge device, is \texttildelow$900ms$, which satisfies the real-time requirement of the framework.
Similar to the approach used in the cloud, each signal-set and its parameters ($W=[S,\omega,\beta]$) are tracked using a lightweight algorithm, which estimates the area ($A$) between the two signals at each time-step.
Signal-sets that do not satisfy the area threshold are removed from the list of signals that are tracked ($F$) in each iteration. 
When the number of signals in this list drops below the signal tracking threshold $H$, the patient's EEG signal for the current time step is transmitted to the cloud to obtain a new $T$ that can be used for tracking the signals once-again.
This procedure is done in the background, i.e., the signal tracking at the edge is still ongoing to provide anomaly prediction probabilities in real-time, while the cloud is used to search for a new signal correlation set $T$.

\setcounter{figure}{7}
\begin{figure}[h]
    \centering
    \captionsetup{singlelinecheck=false}
    \includegraphics[width = \linewidth]{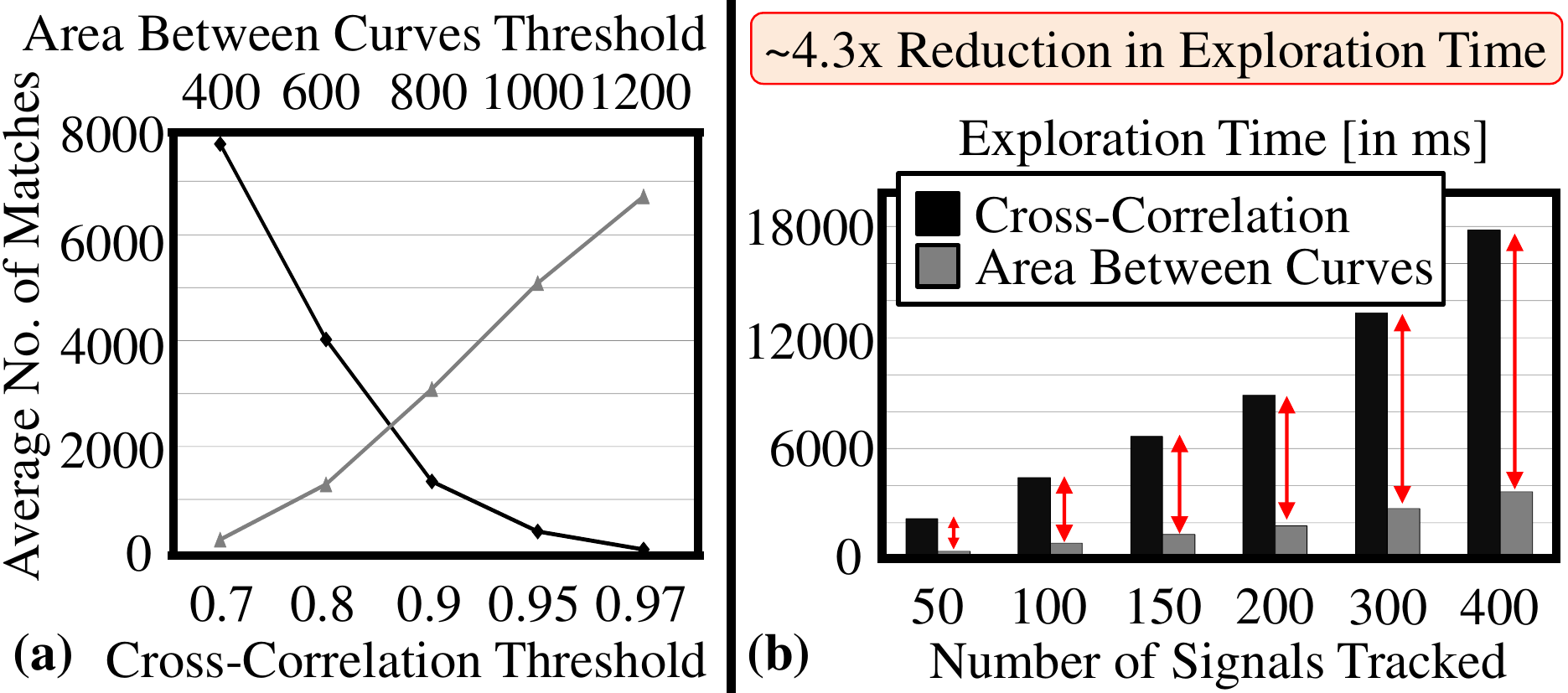}
    \caption{\textbf{Comparison between the cross-correlation approach and the ``area between curves'' approach.}}
    \label{fig:EdgeTracking}
\end{figure}

Figure~\ref{fig:TimingDiagram} presents an overview of the timing analysis of the EMAP framework. 
After the sampling is completed at instance \texttt{a}, the data is filtered and transmitted to the cloud for the mega-database search, which incurs an initial overhead of \texttildelow$3$ seconds.
After the search is complete, the top-$100$ signals are transmitted to the edge device, at instance \texttt{c}, for real-time tracking and probability estimation.
In each iteration $I_N$, we use the proposed lightweight signal tracking algorithm to remove dissimilar signals and to estimate the probability of an anomaly.
If the number of signals being tracked falls below the pre-determined threshold, the previous set of sampled signals is transmitted to the cloud, at instance \texttt{e}, for the \textit{MDB Search}.
The signals are still being tracked in real-time at the edge, i.e., the \textit{MDB Search} is completed at the cloud while doing real-time signal tracking at the edge in parallel.
The same tracking procedure is repeated at the edge with a new set of top-$100$ signals at instance \texttt{h}.
Based on our experiments, we have determined that the sampled EEG signals need to be transmitted to the cloud every five iterations, i.e., after $5$ seconds of edge-processing.

%% file: sections/section6.tex
\section{Experimental Results \& Discussion}
\label{sec:Results}

\subsection{Experimental Setup}
We implemented the proposed EMAP framework by considering an Intel Core i7-7700HQ microprocessor with 16GB of DDR4 RAM and a 128 GB SSD running the Linux Ubuntu 18.04.3 LTS operating system as the cloud, and a Raspberry Pi B+ with 16GB extended memory as the edge node.
The complete framework was implemented in the Python~$3$ programming language with the help of the \textit{scipy}, \textit{sklearn}, \textit{spyedflib}, and \textit{pymongo} libraries.
We used the \textit{MongoDB} framework to implement the $MDB$, to systematically store and access signals.
The $MDB$ was constructed using the signals obtained from the datasets presented in~\cite{PhysioNet}\cite{harati2014tuh}\cite{Dua:2019}\cite{brunner2014bnci}\cite{zwolinski2010open}.

\begin{table*}[t]
    \centering
    \caption{\protect\centering\textbf{Average prediction accuracy of EMAP for three different neurological disorders, compared with he state-of-the-art prediction and detection techniques.}{ \normalfont *\textcolor{red}{N.A.} $\rightarrow$ technique not applicable for the given scenario.}}
    \begin{tabular}{L{1cm}|C{0.66cm}|C{0.66cm}|C{0.66cm}|C{0.66cm}|C{0.66cm}|C{0.66cm}|C{0.66cm}|C{0.66cm}|C{0.66cm}|C{0.66cm}|} 
        \cline{2-11}
        & \multicolumn{5}{c|}{EMAP} & \multicolumn{2}{c|}{SoA - \textit{Prediction}} & \multicolumn{3}{c|}{SoA - \textit{Detection}} \\\cline{2-11}
        & \textbf{B1} & \textbf{B2} & \textbf{B3} & \textbf{B4} & \textbf{B5} & \cite{hosseini2016cloud} & \cite{samie2018highly} & \cite{burrello2019laelaps} & \cite{pascual2019self} & \cite{zhang2015seizure}\\ \hline
        \multicolumn{1}{|l|}{Seizure} & 0.95 & 0.94 & 0.95 & 0.97 & 0.94 & 0.94 & 0.93 & 0.86 & 0.93 & 0.99\\ \hline
        \multicolumn{1}{|l|}{Encephalopathy} & 0.67 & 0.76 & 0.74 & 0.76 & 0.72 & \textcolor{red}{N.A.} & \textcolor{red}{N.A.} & \textcolor{red}{N.A.} & \textcolor{red}{N.A.} & \textcolor{red}{N.A.}\\ \hline
        \multicolumn{1}{|l|}{Stroke} & 0.74 & 0.85 & 0.80 & 0.78 & 0.77 & \textcolor{red}{N.A.} & \textcolor{red}{N.A.} & \textcolor{red}{N.A.} & \textcolor{red}{N.A.} & \textcolor{red}{N.A.}\\ \hline
    \end{tabular}
    \label{table1}
\end{table*}

\subsection{Prediction Accuracy Analysis}
We have already illustrated that the framework's parameters and stages are configured in a manner so as to achieve real-time anomaly predictions.
Therefore, in this section, we will study the accuracy of predictions and the ability of the framework to predict three different neurological disorders.
For the following experiments, we have randomly constructed 5 batches of 20 input signals each to estimate the accuracy of predicting each anomaly that we have considered.
The prediction results presented are for two sequential \textit{cloud calls}, i.e., after transmitting the input signal to the cloud twice after the signal tracking threshold $H$ is violated.

\setcounter{figure}{9}
\begin{figure}[b]
    \centering
    \captionsetup{singlelinecheck=false}
    \includegraphics[width = \linewidth]{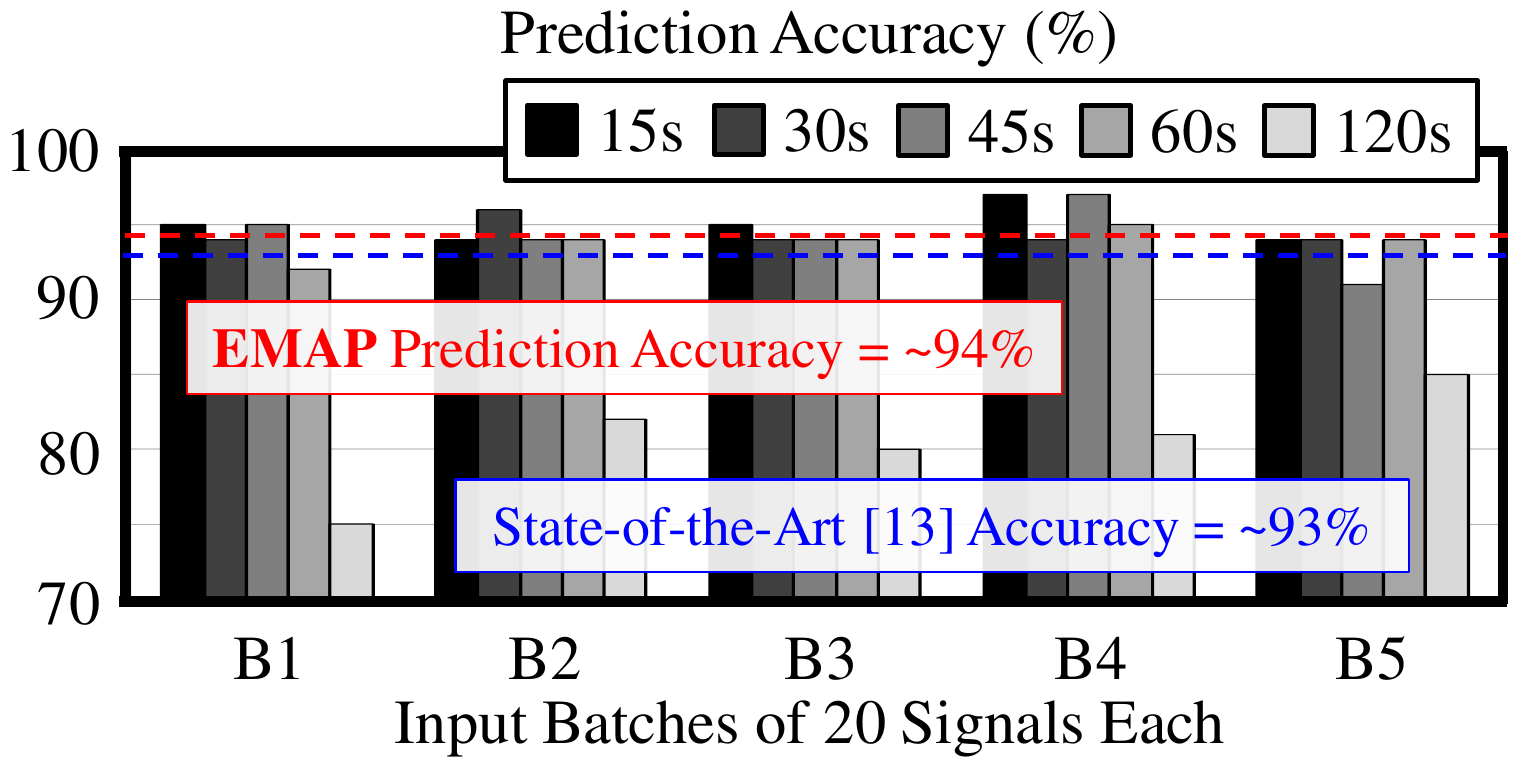}
    \caption{\textbf{Prediction accuracy analysis of EMAP for anomaly 1 (seizure) at 15, 30, 45, 60 and 120 second intervals before the occurrence of a seizure, and its comparison with the state-of-the-art IoT-based seizure prediction technique~\cite{samie2018highly}}.}
    \label{fig:PredictionSeizure}
\end{figure}

\textit{Seizures} are one of the most common neurological disorders in the world, and therefore, is one of the diseases that has been widely studied as a research challenge.
Fig.~\ref{fig:PredictionSeizure} presents the prediction accuracy results of the \textit framework at various time intervals before the occurrence of the seizure.
We have achieved a maximum prediction accuracy of $97\%$ and an average prediction accuracy of $94\%$ in real-time.
Each time-step of the input signal is compared with the set of correlated signals to estimate the anomaly probability, which if increasing is classified as an anomaly.
Whereas the state-of-the-art technique~\cite{samie2018highly}, on average, achieves a prediction accuracy of $93\%$.
Furthermore, this technique is highly specific and can only be used to predict the occurrence of \textit{seizures}, whereas the EMAP framework can be used to predict multiple different brain-anomalies.

Next, we evaluate the proposed EMAP framework for other anomalies, such as \textit{encephalopathy} (Anomaly 2) and \textit{stroke} (Anomaly 3), the results of which are presented in Table~\ref{table1}.
Due to the unavailability of similar highly annotated datasets for these two anomalies, i.e., the preset and onset of anomaly progression, for the following two cases, we have annotated the complete signal as an anomaly.
We have achieved an average prediction accuracy of $73\%$ and $79\%$, respectively for encephalopathy and strokes. This reduction in prediction accuracy is attributed to the unavailability of a substantially-labeled dataset such as the ones available for the seizure.
Furthermore, since the proposed algorithm focuses on maximizing the sensitivity to anomalies and classifies near-threshold anomaly probability increases as anomalous, the classification accuracy of the normal signal is reduced, i.e., the average percentage of false-positives is \texttildelow$15\%$, which is a limitation of the EMAP framework.

Finally, we evaluate the loss in accuracy of deploying the proposed signal cross-correlation search (Algorithm~\ref{Algo1}) in the cloud instead of the time-consuming exhaustive cross-signal.
We evaluate the average signal cross-correlation of the top-$100$ signals with respect to the input for $100$ different normal and anomalous input signals.
The results of these experiments are illustrated in Fig.~\ref{fig:AccuracyAnalysis}.
As can be observed, \textit{the average cross-correlation of the proposed approach is very close to the average cross-correlation of the signals obtained using the exhaustive cross-correlation technique}.
Therefore, the loss in accuracy is almost non-existent and indistinguishable due to the substantially large and highly redundant data-set that we use.
However, due to the sliding window technique deployed in the proposed approach, the top-$100$ signals selected are very diverse, i.e., typically, the top-$100$ signals exhibit high cross-correlation to the input, but can also exhibit very low cross-correlation in certain scenarios, as illustrated in the figure.

\setcounter{figure}{10}
\begin{figure}[h]
    \centering
    \captionsetup{singlelinecheck=false}
    \includegraphics[width = \linewidth]{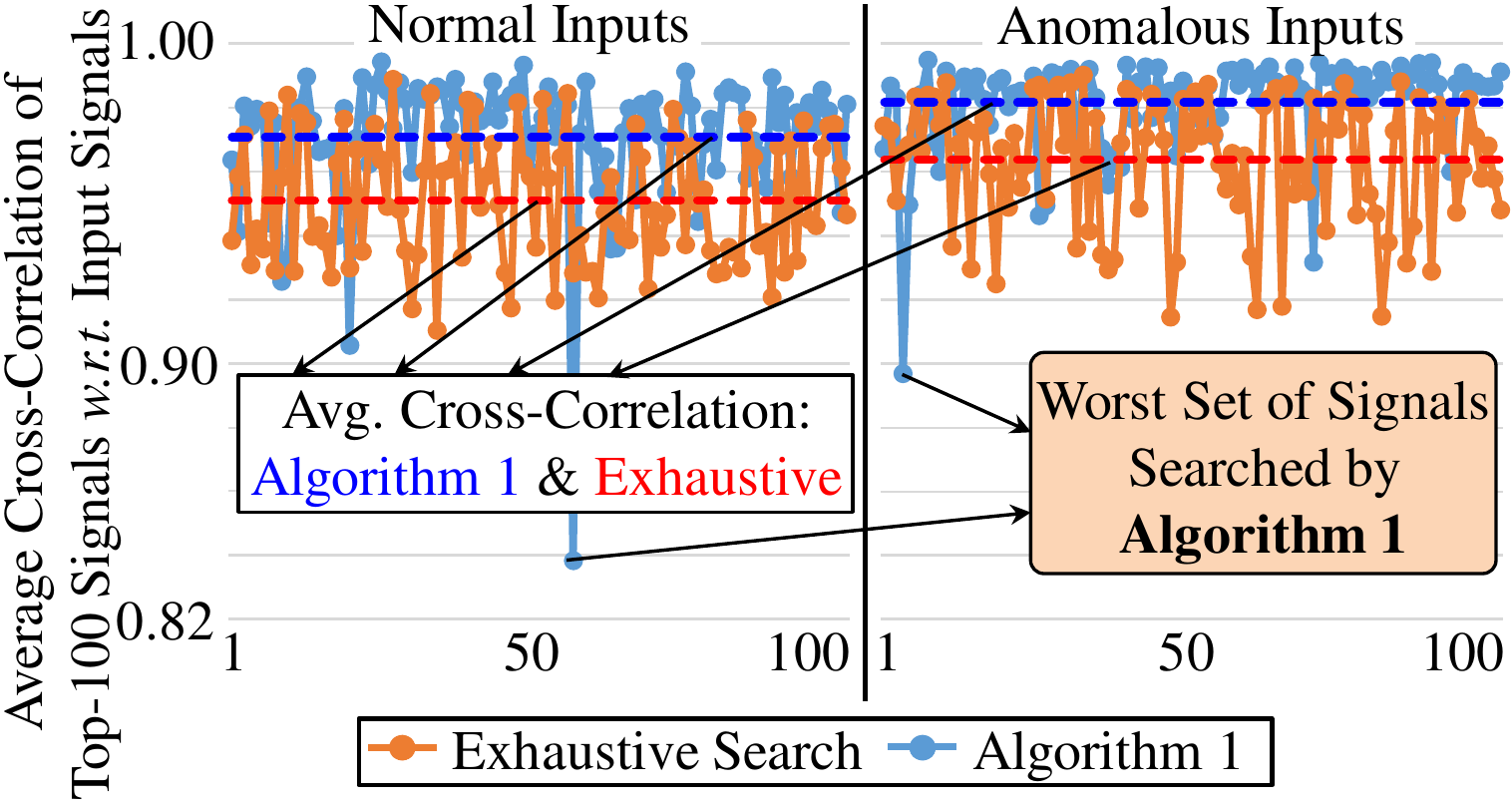}
    \caption{\textbf{Accuracy Evaluation of the Proposed Algorithm~\ref{Algo1} Compared to the Exhaustive Cross-Correlation Search.}}
    \label{fig:AccuracyAnalysis}
\end{figure}

%% file: sections/section7.tex
\section{Conclusion}
\label{sec:Conclusion}

In this work, we presented EMAP, a cloud-edge hybrid framework that is beneficial for continuously monitoring EEG signals and to estimate the probability of occurrence of an anomaly in real-time.
The framework is composed of three key stages, namely,
\begin{inlinelist}
    \item \textit{Signal Acquisition}, which is responsible for collecting, filtering, and transmitting the EEG data to the cloud;
    \item \textit{Cloud Search}, where the input signal is cross-correlated with all the signals in the $MDB$, which is a construction of multiple openly accessible EEG dataset, to determine the top-$100$ signals with maximum similarity to the input signal; and
    \item \textit{Edge Tracking}, where the subsequent EEG signal samples are used to eliminate the dissimilar signals and predict the occurrence of an anomaly.
\end{inlinelist}
Using the proposed, we have achieved a prediction accuracy of $94\%$, $73\%$, and $79\%$ for three different anomalies, namely, seizures, encephalopathy, and strokes, respectively.
The EMAP framework has been made open-source at \textcolor{blue}{\url{https://emap.sourceforge.io}}, to ensure ease of adoption and reproducibility.